\documentclass[fleqn]{NMDofRMF}
\begin{document}

\ensubject{subject}

%%%%%%%%%%%%%%%%%%%%%%%%%%%%%%%%%%%%%%%%%%%%%%%%%%%%%%%
%%% Authors do not modify the information below
%%% ????????????????
%%% ??????????, ????????????{}, ???????????????????
%Letter to the Editor??Article%??????
\ArticleType{Article}%??Article
\SpecialTopic{SPECIAL TOPIC: }%???????
\Year{2021}
\Month{March}
%\Vol{63}
%\No{1}
%\DOI{??}
%\ArtNo{000000}
%\ReceiveDate{January 11, 2020}
%\AcceptDate{April 6, 2020}
%\OnlineDate{January 1, 2016}
%%%%%%%%%%%%%%%%%%%%%%%%%%%%%%%%%%%%%%%%%%%%%%%%%%%%%%%

%%% title: ????
%%%   \title{title}{title for citation}
\title{Nucleon momentum distribution of $^{56}\text{Fe}$ from the axially deformed relativistic mean-field model with nucleon--nucleon correlations}{Nucleon momentum distribution of $^{56}\text{Fe}$ from the axially deformed relativistic mean-field model with nucleon--nucleon correlations}

%%% Corresponding author: ???????
%%%   \author[number]{Full name}{{email@xxx.com}}
%%% General author: ???????
%%%   \author[number]{Full name}{}
\author[1]{Xuezhi Wang}{}%
\author[1]{Qinglin Niu}{}
\author[2]{Jinjuan Zhang}{}%
\author[3]{Mengjiao Lyu}{}
\author[1]{Jian Liu}{{liujian@upc.edu.cn}}
\author[4]{\\Chang Xu}{}
\author[5]{Zhongzhou Ren}{}

%%% Author information for page head. ??????????
%%% ??????????????, ??????????author???
\AuthorMark{Xuezhi Wang}%\authorcr????????

%%% Authors for citation. ?????????????????
%%% ??????????????, ??????????author???
\AuthorCitation{Xuezhi Wang, Qinglin Niu, Mengjiao Lyu, et al}

%%% Address. ???
%%%   \address[number]{Address, City {\rm Postcode}, Country}
\address[1]{College of Science, China University of Petroleum (East China), Qingdao 266580, China}
\address[2]{College of Electronic and Information Engineering, Shandong University of Science and Technology, Qingdao 266590, China
}
\address[3]{ College of Science, Nanjing University of Aeronautics and Astronautics, Nanjing 210016, China}
\address[4]{Department of Physics, Nanjing University, Nanjing 210093, China
}
\address[5]{School of Physics Science and Engineering, Tongji University, Shanghai 200092, China}

%\contributions{}%????????

%%% Abstract. ??
\abstract{Nucleon momentum distribution (NMD), particularly its high-momentum components, is essential for understanding the nucleon--nucleon ($ NN $) correlations in nuclei.  Herein, we develop the studies of NMD of $^{56}\text{Fe}$ from the axially deformed relativistic mean-field (RMF) model.  Moreover, we introduce the effects of $ NN $ correlation into the RMF model from  phenomenological models based on deuteron and nuclear matter. For the region  $ k<k_{\text{F}}  $, the effects of  deformation on the NMD of  the RMF model are investigated using the total and single-particle NMDs. For the region $ k>k_{\text{F}} $, the high-momentum components of the RMF model are modified by the  effects of $ NN $ correlation, which agree with the  experimental data. Comparing the NMD of relativistic and non-relativistic mean-field models, the relativistic effects on nuclear structures in momentum space are analyzed.  Finally, by analogizing the tensor correlations in deuteron and Jastrow-type correlations in nuclear matter, the behaviors and contributions of $ NN $ correlations in $^{56}\text{Fe}$ are further analyzed, which helps clarify the effects of the tensor force on the NMD of heavy nuclei.}%??

%%% Keywords. ?????


\maketitle

%\tableofcontents%?????

%%%%%%%%%%%%%%%%%%%%%%%%%%%%%%%%%%%%%%%%%%%%%%%%%%%%%%%
%%% The main text. ???????
%???????????????????\cref{fig1}
%\twocolumn\onecolumn
%%%%%%%%%%%%%%%%%%%%%%%%%%%%%%%%%%%%%%%%%%%%%%%%%%%%%%%
\begin{multicols}{2}
\section{Introduction}\label{section1}

As a fundamental problem in nuclear physics, nucleon momentum distribution (NMD) has attracted research interest for a long time \cite{Pandharipande1997,Guo2017,antonov2012}. On the one hand, NMD can directly reflect the strong interaction in nuclei \cite{ciofi2015}. The high-momentum components of NMD provide an important window for understanding the nucleon--nucleon short-range  correlation ($ NN $-SRC) \cite{pieper2001,Yong2017,Xu2020,Shang2020,ohen2014}. On the other hand, previous    studies on NMD indicated that many exotic nuclear  structures, such as \Authorfootnote\noindent the skin \cite{Cai2016} and  cluster \cite{Colle2015,Li2020}, exist   in  momentum space. For instance, proton skins in momentum   space and neutron skins in coordinate space coexist in heavy nuclei \cite{Cai2016}. These studies are extremely important for a complete understanding of exotic nuclear structures.\par
Remarkable experimental progress has been achieved in deriving and analyzing the NMD \cite{Kelly2002,Cao2011}. A valid method for investigating the NMD is quasielastic electron scattering \cite{Sick1980,Hen2017}. The momentum distributions of nuclei $ ^{3,4} $He, $ ^{12} $C, and $ ^{56} $Fe are extracted from  $y$-scaling analyses on inclusive electron scattering experiments $ A(e,e') $ \cite{Day1990,Fomin2012,ciofi1991,ciofi1996}. Moreover, relative abundances of the $ NN $-SRC  $ pn $,  $ nn $, and $ pp $ pairs in nuclei from $ ^{12} $C to $ ^{208} $Pb are obtained in  exclusive electron scattering experiments $ A(e,e'pp) $ and $ A(e,e'pn) $ \cite{Tang2003,Benmokhtar2005,Piasetzky2006,Shneor2007,Subedi2008,Dai2017,Korover2014}. Based on various experimental evidence,  NMDs have been reported to have two distinct regions: the mean-field region $ \left( k<k_{\text{F}}\right)  $, which contains approximately  80\% of the nucleons, and the high-momentum region $ \left( k>k_{\text{F}}\right)  $, which only contains approximately 20\% \cite{ohen2014,Tang2003} of the nucleons but contains approximately 70\% \cite{Polls1994} of the nuclear kinetic energy. In the mean-field region, the many-body dynamics cause single nucleons to move under the influence of an effective potential, whereas in the high-momentum component, the nucleons are in the form of the $ NN $-SRC pairs, particularly the $ pn $-SRC  pairs induced by the tensor force \cite{Alvioli2008,Collaboration2018,Collaboration2019,Duer2019}. \par
 For  theoretical studies on finite nuclear systems, the $ ab \ initio $ method \cite{Pieper2001mk,Liu2014,Hagen2015,Sun2016,Wan2020} is a fundamental approach that describes the nucleon--nucleon $ \left( \text{$ NN $}\right)  $ correlations by solving  dynamical equations. Using  effective pair-based generalized contact formalism and $ ab \ initio $ quantum Monte Carlo calculations, the high-momentum components of nuclei from deuteron to $ ^{40}$Ca have been studied \cite{Wiringa2014,Lonardoni2018,Cruz-Torres2020}. The physical origins of the high-momentum components, such as the contribution of the correlations induced by tensor attraction and short-range repulsion, are further clarified. In Ref.  \cite{Wiringa2014}, the authors used  realistic  AV18+UX   potentials to generate accurate $ab \ initio$ wave functions for light nuclei and proved that tensor correlations play a dominant role in the $ 1.5 \  \text{fm} ^{-1}\sim3.0\ \text{fm} ^{-1}  $ range of NMDs. Decomposition analyses  were  performed for the NMDs of the $ ^{4} $He nucleus in Ref. \cite{Lyu2020} via $ ab \ initio $ calculations using $  \text{AV}8' $ bare interaction, which showed that the tensor correlation and the short-range repulsion dominated around $ 2 \ \text{fm} ^{-1} $  and $ 4 \ \text{fm} ^{-1} $, respectively.  For heavier nuclei, $ab \ initio$ calculations of nuclear structures are still limited, but rapid progress is underway \cite{Carlson2015}. \par
 Instead of the $ ab \ initio $ method, the many-body model is another effective method for  studying  heavy nuclei, which can accurately describe and predict nuclear structures. A successful many-body theoretical method is the self-consistent mean-field model \cite{Bender2003,Liu2013,Wang2020,Meng2016}. This method describes the motions of the nucleons using an overall mean field, and has calculated a range of nuclear properties with remarkable accuracy. In Ref. \cite{Deguerra1991}, the authors calculated the NMDs of Nd isotopes using the non-relativistic Skyrme Hartree-Fock (SHF) model. The SHF model results were expected to be reliable up to Fermi momentum $ k_{\text{F}} $. However, because of the absence of $ NN $ correlations in the mean-field framework, the NMDs contradicted the experimental data in the high-momentum region $ k>k_{\text{F}} $. \par
 Two methods are available for importing the effects of the $ NN $ correlations into the mean field. One such method is the light-front dynamics (LFD) method \cite{Antonov2002}. In the framework of the LFD method,  the deuteron wave function is expressed using the six scalar functions calculated via the relativistic one-boson-exchange model \cite{Carbonell1995}. For finite nuclei, the effects of $ NN $ correlation on NMD are introduced through the rescaling of the high-momentum components of deuteron based on the natural orbital representation. The other method is the local density approximation (LDA) method \cite{Stringari1990}. This method uses the $ NN $ correlations in the nuclear matter to modify the occupation probability predicted by the Fermi gas model. For finite nuclei, the correlation effects are generated from the uniform nuclear matter calculations at different densities \cite{Flynn1984}. Combining the non-relativistic many-body models with the LFD and LDA methods, the calculations can successfully describe the high-momentum components of NMD in heavy nuclei \cite{Antonov2005,Antonov2006,Gaidarov2009}. However, further research is required to analyze the physical origin of the $ NN $ correlations, such as the impacts of the tensor force and short-range repulsion. \par
 Compared with the non-relativistic many-body models, the relativistic mean-field (RMF) model \cite{Ring1996,zhou2004,Wang2019,Liu2017,Zhao2020,Long2020} is another mean-field approach for calculating the nuclear bulk properties of $ r $-space \cite{Li2019,Meng2019}, such as the nuclear masses \cite{Hua2012}, charge radii \cite{Mayun2020}, density distributions \cite{Liu2013prc,Ren2017,Liang2018,Liu2019,Ren2020}, and \textit{et al}. . However, there is inadequate research for the NMD of the RMF model. Although  the RMF and SHF models are both built in the mean-field framework, notable discrepancies exist in the central potentials and single-particle wave functions between these two models \cite{Ebran2012}.  Therefore, analyzing the NMD based on the RMF model and comparing the calculations with previous results would be interesting and significant.\par
 The main purpose of this work is to study the NMD of the heavy nuclei within the framework of the deformed RMF model incorporating the nucleon--nucleon correlations.  We choose $^{56}\text{Fe}$ as the candidate nucleus in the calculations because it has the existing data for the NMD extracted from the $y$-scaling analyses of inclusive electron scattering off nuclei \cite{ciofi1991,ciofi1996}.  First, we investigate the total and single-particle NMDs to discuss the influences of deformation on RMF calculations in momentum space.  Next, the comparisons on NMD between the RMF and non-relativistic SHF models are  performed to analyze the relativistic effects on the nucleon wave function in momentum space.  Finally, the LFD and LDA methods are used to introduce the $ NN $ correlations to correct the high-momentum components of NMD in $^{56}\text{Fe}$. The theoretical NMDs are also compared with the experimental data to validate the integration between the deformed RMF model and  $ NN $ correlations. Finally, we decompose the $ NN $ correlations into different physical terms for the LFD and LDA methods, respectively, and further perform  fine-grained analyses for their behaviors and contributions to the total NMD in $^{56}\text{Fe}$. \par
 The article is organized as follows: \cref{sec:2} describes the NMD of the axially deformed RMF model and the amendments for the high-momentum components using the LFD and LDA methods. \cref{sec:3} presents numerical results and discussions. Finally, in \cref{sec:4} we draw the main conclusions of this work.

\section{Theoretical framework}\label{sec:2}
\subsection{NMD from the deformed RMF model}\label{sec:2.1}
In this subsection, we discuss the NMD of the axially deformed RMF model. In the RMF model, the starting point is the Lagrangian density \cite{Todd-Rutel2005}
\begin{align}
	\begin{aligned}
		\mathcal{L}=& \bar{\psi}\bigg\{\gamma^{\mu}\left[i \partial \mu-g_{\omega} \omega_{\mu}-\frac{g_{\rho}}{2} \rho_{\mu}^{a} \tau^{a}-\frac{e}{2}\left(1+\tau^{3}\right) A_{\mu}\right]\\
		&-\left(M-g_{s} \sigma\right)\bigg\} \psi \\
		&+\frac{1}{2} \partial^{\mu} \sigma \partial_{\mu} \sigma-\frac{1}{2} m_{\sigma}^{2} \sigma^{2}-\frac{\kappa}{3 !}\left(g_{s} \sigma\right)^{3}-\frac{\lambda}{4 !}\left(g_{s} \sigma\right)^{4} \\
		&-\frac{1}{4} \varOmega^{\mu v} \varOmega_{\mu v}+\frac{1}{2} m_{\omega}^{2} \omega^{\mu} \omega_{\mu}+\frac{\xi}{4 !}\left(g_{\omega}^{2} \omega^{\mu} \omega_{\mu}\right)^{2} \\
		&-\frac{1}{4} \vec{R}^{\mu v} \cdot \vec{R}_{\mu v}+\frac{1}{2} m_{\rho}^{2} \vec{\rho}^{\mu} \cdot \vec{\rho}_{\mu} \\
		&-\frac{1}{4} F^{\mu v} F_{\mu \nu}+\Lambda_{\mathrm{v}}\left(g_{\rho}^{2} \vec{\rho}^{\mu} \vec{\rho}_{\mu}\right)\left(g_{\omega}^{2} \omega^{\mu} \omega_{\mu}\right).
	\end{aligned}\label{eq:1}
\end{align} \par
Based on the variational principle, Dirac equations for the nucleons and  Klein--Gordon equations for the mesons can be deduced. For axisymmetric deformed shapes, the Dirac spinors can be expressed as follows \cite{Meng2016}:
\begin{align}
	\begin{aligned}
		\psi _i\left( \vec{r},t,\sigma \right) =&\left( \begin{array}{c}
			f_i\left( \vec{r} \right)\\
			ig_i\left( \vec{r} \right)\\
		\end{array} \right) \chi _{t_i}\left( t \right) \chi _{\varSigma}\left( \sigma \right)\\
		=&\frac{1}{\sqrt{2\pi}}\left( \begin{array}{c}
			f_{i}^{+}\left( z,r_{\bot} \right) e^{i\left( \varOmega _i-1/2 \right) \varphi}\\
			f_{i}^{-}\left( z,r_{\bot} \right) e^{i\left( \varOmega _i+1/2 \right) \varphi}\\
			ig_{i}^{+}\left( z,r_{\bot} \right) e^{i\left( \varOmega _i-1/2 \right) \varphi}\\
			ig_{i}^{-}\left( z,r_{\bot} \right) e^{i\left( \varOmega _i+1/2 \right) \varphi}\\
		\end{array} \right) \chi _{t_i}\left( t \right) \chi _{\varSigma}\left( \sigma \right),
	\end{aligned}\label{eq:2}
\end{align}
where $\varOmega _i$ is the eigenvalue of the projection of the single-particle angular momentum on the symmetry axis. Two spinors $f_i$ and $g_i$ can be expanded by the eigenfunctions $\Phi _{\alpha}$ of an axially symmetric deformed harmonic-oscillator potential in cylindrical coordinates
\begin{align}
\begin{aligned}
	f_i\left( \vec{r},\sigma ,t \right) &=\sum_{\alpha}^{\alpha _{\max}}{f_{\alpha}^{\left( i \right)}}\Phi _{\alpha}\left( \vec{r},\sigma \right) \chi _{t_i}\left( t \right),\ \\
	g_i\left( \vec{r},\sigma ,t \right) &=\sum_{\tilde{\alpha}}^{\tilde{\alpha}_{\max}}{g_{\tilde{\alpha}}^{\left( i \right)}}\Phi _{\tilde{\alpha}}\left( \vec{r},\sigma \right) \chi _{t_i}\left( t \right),
\end{aligned}\label{eq:3}
\end{align}
with the quantum numbers $ \alpha =\left\{ n_z,n_r,\varLambda ,\varSigma \right\} $.\par
Using \cref{eq:2,eq:3}, we further deduce the Dirac spinors in momentum space  through Fourier transformation
\begin{align}
\begin{aligned}
	\tilde{f}_i\left( \vec{k},\sigma ,t \right) &=\sum_{\alpha}^{\alpha _{\max}}{f_{\alpha}^{\left( i \right)}}\Phi _{\alpha}\left( \vec{k},\sigma \right) \chi _{t_i}\left( t \right),\  \\
	\tilde{g}_i\left( \vec{k},\sigma ,t \right) &=\sum_{\tilde{\alpha}}^{\tilde{\alpha}_{\max}}{g_{\tilde{\alpha}}^{\left( i \right)}}\Phi _{\tilde{\alpha}}\left( \vec{k},\sigma \right) \chi _{t_i}\left( t \right),
\end{aligned}
\label{eq:4}
\end{align}
where
\begin{align}
	\varPhi _{\alpha}\left( \vec{k},\sigma \right) =\frac{1}{\left( 2\pi \right) ^{3/2}}\int{\text{d}}\vec{r} \text{e}^{-i\vec{k}\cdot \vec{r}}\varPhi _{\alpha}\left( \vec{r},\sigma \right).\label{eq:5}
\end{align}
Introducing the corresponding oscillator length parameters $ b_z=\sqrt{\frac{\hbar }{M\omega _z}}$ and $ b_{\bot}=\sqrt{\frac{\hbar }{M\omega _{\bot}}} $, the eigenfunctions of the deformed harmonic oscillator are represented by separate variable functions
\begin{align}
	\begin{aligned}
		\Phi _{\alpha}\left( \vec{k},\sigma \right) &=\phi _{n_z}\left( k_z \right) \phi _{n_r}^{\varLambda}\left( k_{\bot} \right) \frac{1}{\sqrt{2\pi}}e^{i\varLambda \varphi}\chi _{\varSigma}\left( \sigma \right),\\
	\end{aligned}\label{eq:6}
\end{align}
where the functions of $  z $ and $ r $ directions are given by
\begin{align}
	\begin{aligned}
		\phi _{n_z}\left( k_z \right) &=\sqrt{\frac{b_z}{\sqrt{\pi}2^{n_z}n_z!}}H_{n_z}\left( k_{\xi} \right) e^{-k_{\xi}^2/2},\\
		\phi _{n_r}^{\varLambda}\left( k_{\bot} \right) &=\sqrt{\frac{n_r!}{\left( n_r+\varLambda \right) !}}b_{\bot}\sqrt{2}k_{\eta}^{\varLambda /2}L_{n_r}^{\varLambda}\left( k_{\eta} \right) e^{-k_{\eta}/2}\\
	\end{aligned}\label{eq:7}
\end{align}
with two dimensionless variables $k_{\xi} =k_zb_z$ and $ k_{\eta} =k_{\bot}^{2}b_{\bot}^{2}$. The polynomials  $ H_{n_z}\left( k_{\xi} \right) $ and $L_{n_r}^{\varLambda}\left( k_{\eta} \right) $ are Hermite polynomials and associated Laguerre polynomials, respectively. Thus, we express the densities in momentum space for state $ i $, namely the single-particle  NMD can be expressed as follows: 
\begin{align}
	\begin{aligned}
		n_i\left( k_{\bot},k_z \right) =&\frac{1}{\pi}\sum_{\alpha \alpha'}{\left( -i \right)^{N+N'}\left(f_{\alpha}^{\left(i\right)}f_{\alpha'}^{\left(i\right)}+g_{\alpha}^{\left(i\right)}g_{\alpha'}^{\left(i\right)}\right)}\delta_{\varLambda\varLambda^{'}}\delta _{\varSigma\varSigma^{'}}\\
		&\times\phi _{n_z}\left( k_\xi \right) \phi _{n_z'}\left( k_\xi \right) \phi _{n_r}^{\varLambda}\left( k_\eta \right) \phi _{n_r'}^{\varLambda}\left( k_\eta \right),
	\end{aligned}\label{eq:8}
\end{align}
where $ N $ is the major quantum number, $ N=n_z+2n_r+\varLambda $, and $\left( -i \right) ^{N+N'}$ is the product of the two phase factors $\left( -i \right) ^{N}$ and $\left( -i \right) ^{N'}$ from  Fourier transformation. Notably, for negative-parity states that have odd N, the above-mentioned phase factor can be changed as $\left( -i \right) ^{N-1}$ \cite{Deguerra1991}. Correspondingly, the total NMD is defined as follows:
\begin{align}
	n\left( k_{\bot},k_z \right) =\sum_i{2\lambda _i}n_i\left( k_{\bot},k_z \right).\label{eq:9}
\end{align}
\par
After  $n\left( k_{\bot},k_z \right) $ are obtained, we further expand the total NMD using the Legendre function
\begin{align}
	n\left( k_{\bot},k_z \right) =n\left( k\cos \theta _k,k\sin \theta _k \right) =\sum_{\lambda}{n_{\lambda}}\left( k \right) P_{\lambda}\left( \cos \theta _k \right)\label{eq:10}
\end{align}
with multipole components $ \lambda $
\begin{align}
	\begin{aligned}
		n_{\lambda}\left(k \right)\!=\!\frac{2\lambda\!+\!1}{2}\!\int_{-1}^1\!{P_{\lambda}}\left( \cos \theta _k \right) n\left( k\cos \theta _k,k\sin \theta _k \right) \text{d}\left( \cos \theta _k \right).
	\end{aligned}
	\label{eq:11}
\end{align}
\subsection{Effects of the $ NN $ correlations on NMD}\label{sec:2.2}
Because of the lack of $ NN $ correlations in nuclear wave functions, self-consistent mean-field calculations cannot explain the high-momentum components of the momentum distributions \cite{Deguerra1991,antonov2012,ohen2014}. In this subsection, the effects of $ NN $ correlations on the NMD in heavy nuclei calculated from the deformed RMF model are introduced using two phenomenological methods---LFD and LDA. In these methods,  NMDs can be generally expressed as follows:
 \begin{align}
 	n_{\tau}\left( k \right) =n_{\tau}^{RMF}\left( k \right) +\delta n_{\tau}\left( k \right), \label{eq:12}
 \end{align}
where $ n_{\tau}^{RMF}\left( k \right) $  is the RMF contribution, and $  \delta n_{\tau}\left( k \right) $  embodies the corrections of the $ NN $ correlations; the subscript $ \tau $ represents the proton or neutron. The high momentum components induced by the $ NN $ correlation, in principle, should be included self-consistently in the calculations; however, this is  difficult within the framework of the present mean-field model. An approach similar to \cref{eq:12} was used in the simulations of heavy ion  collisions \cite{Hui2016,Wang2017}, wherein the authors introduced the $ NN $ correlations into the free Fermi gas model and found that the high-momentum components could  increase  both high-energy free proton and neutron emission.\par
 \subsubsection{The light-front dynamics method}\label{sec:2.2.1}
  The light-front dynamics (LFD) method is an effective method employing the high-momentum components of the deuteron to introduce the $ NN $ correlations. In the framework of the LFD method, the NMD starts from the natural-orbital representation \cite{antonov2012}
  \begin{align}
  	n_{\tau}\left( k \right)=n_{\tau}^{h}\left( k \right) +n_{\tau}^{p}\left( k \right) \label{eq:13}
  \end{align}
with the following normalization condition: $\int{n_{\tau}}\left( k \right) d\vec{k}=\tau $. $ n_{\tau}^{h}\left( k \right) $  and   $ n_{\tau}^{p}\left( k \right)  $ are the hole-state and particle-state contributions, respectively. With  the single-particle NMDs in \cref{eq:10}, the $ n_{\tau}^{h}\left( k \right) $  and  $ n_{\tau}^{p}\left( k \right)  $  can be expressed in the axially deformed RMF model as follows:
  \begin{align}
  		\begin{aligned}
  		n_{\tau}^{h}\left( k \right) =&\frac{1}{\tau}\sum_i^{FL}{2\lambda _i}C\left( k \right) n_i\left( k_{\bot},k_z \right), \\
  		n_{\tau}^{p}\left( k \right) =&\frac{1}{\tau}\sum_{FL}^{i_{max}}{2\lambda _i}C\left( k \right) n_i\left( k_{\bot},k_z \right),
  	\end{aligned} \label{eq:14}
  \end{align}
where the FL denotes the Fermi level and
  \begin{align}
  	C\left( k \right) =\frac{m}{\left( 2\pi \right) ^3\sqrt{k^2+m^2}}.\label{eq:15}	
  \end{align}
  For the normalization condition of \cref{eq:13,eq:14}, we substitute $ n_{\tau}^{h}\left( k \right) $ by
  \begin{align}
  	\tilde{n}_{\tau}^{h}\left( k \right) =\frac{\tau \sum_i^{FL}{2\lambda _i}C\left( k \right) n_i\left( k_{\bot},k_z \right)}{\int{C\left( k \right) n\left( k_{\bot},k_z \right)}d\vec{k}}.\label{eq:16}
  \end{align}\par
In the framework of mean-field theory,  $n^p\left( k \right) $ in \cref{eq:13}, which plays a leading role in the  high-momentum components of NMD, cannot well reproduce the experimental results owing to the absence of $ NN $ correlation contributions. For the expressions of the high-momentum components, it remains challenging in the $ab \ initio$ calculations for nuclear systems to correctly describe the strong $ NN $ correlations in the heavy nuclei. However, previous studies \cite{ciofi1996,Alvioli2013,Hen2017} have shown that the high-momentum components of NMDs are approximately equal for all nuclei. Therefore, in the framework of LFD, the $
 n^p\left( k \right)
 $
   of the $ A>2 $ nuclei can be obtained by rescaling the particle-state contributions of the precise NMD of deuteron  $ n_{d}^{p}\left( k \right) $
  \begin{align}
  	n^p\left( k \right) \cong \ C_An_{d}^{p}\left( k \right),\label{eq:17}
  \end{align}
where  $ C_A $ is the scaling factor for the high-momentum components between deuteron and other nuclei. The NMD of deuteron is determined by three components---$ n_1\left( k\right)  $, $ n_2\left( k\right)  $, and $ n_5\left( k\right)  $---deduced from six LFD wave functions \cite{Carbonell1995}. In the high-momentum components at $k > 2\,\text{fm}^{-1} $, the components $ n_2\left( k\right)  $ and $ n_5\left( k\right) $ have the most crucial contributions \cite{Antonov2002,Antonov2005}. Therefore, the high-momentum components of NMD in the LFD method can be expressed as follows:
  \begin{align}
  n_{\tau}^{p}\left( k \right) \cong \ \tau C_A\left[ n_2\left( k \right) +n_5\left( k \right) \right],\label{eq:18}
  \end{align}\par
In \cref{eq:10}, $  n_2\left( k \right) $ is obtained from the second LFD wave function $ f_2 $, a component of the wave function coincided with the Bonn D-wave in the region $ 0<k<3\ \text{fm}^{-1} $, which can also be regarded as the tensor force contribution in this region. The  $  n_5\left( k \right) $ is obtained from the fifth LFD wave functions $ f_5 $, wherein the main effects of $ f_5 $ focus on the $  k > 2.5\ \text{fm}^{-1} $, and its approximately $  60 \% $ are given by $\pi$-exchange. The scaling factor $ C_A$  is listed  Table 1 based on Ref. \cite{ciofi1996}, where the value is related to $ A $, and the reasonable range for $ A>12 $ nuclei is $  4.0 < C_A <5.0 $. Finally, the results of NMDs from the LFD method can be expressed as follows:
  \begin{align}
  	n_{\tau}\left( k \right) =N_{\tau}\left\{ \tilde{n}_{\tau}^{h}\left( k \right) +\tau C_A\left[ n_2\left( k \right) +n_5\left( k \right) \right] \right\}, \label{eq:19}
  \end{align}
  where
  \begin{align}
  	N_{\tau}&=\left\{ 4\pi \int_0^{\infty}{dk\cdot k^2}\left\{ \tilde{n}_{\tau}^{h}\left( k \right) +\tau C_A\left[ n_2\left( k \right) +n_5\left( k \right) \right] \right\} \right\} ^{-1} \label{eq:20}
  \end{align}
  ensures the normalization.
\subsubsection{The local density approximation}\label{sec:2.2.2}
The local density approximation (LDA) approach is another successful method based on the Fermi gas model \cite{Stringari1990} that combines the contributions of mean-field and $ NN $ correlations. In the LDA approach, the $ NN $ correlations are derived from the corresponding results calculated using uniform nuclear matter at different densities and are insensitive to the surface and shell effects \cite{Flynn1984}. The contributions of the $ NN $ correlation $\delta n_{\tau}\left( k \right) $ in \cref{eq:12} can be expressed as follows: 
\begin{align}
	\delta n_\tau\left( k \right) =\frac{1}{4\pi ^3}\int{\delta}v\left[ k_{\text{F}}^{\tau}\left( r \right) ,k \right] d\vec{r}, \label{eq:21}
\end{align}
where $ \delta v\left[k_{\text{F}}^{\tau}\left( r \right),k \right]  $  represents the effects of the $ NN $ correlations and the $ k_{\text{F}}^{\tau}\left( r \right) $  is the local Fermi momentum, $
k_{\text{F}}^{\tau}\left( r \right) =\sqrt[3]{3\pi ^2\rho _{\tau}\left( r \right)} $. By the definition of Fermi momentum in the Fermi gas model, one has  $\int{\delta v\left[k_{\text{F}}^{\tau}\left(r\right),k\right]}d\vec{k}=0 $. \par
Based on the results of the lowest-order cluster (LOC) approximation  \cite{Flynn1984}, the additional item   $ \delta v\left[k_{\text{F}}^{\tau}\left( r \right),k \right]  $, is given as
\begin{align}
	\begin{aligned}
		\delta v\left[ k_{\text{F}}^{\tau}\left( r \right) ,k \right] =&\left[ Y\left( k,8 \right) -\kappa _{\text{dir}} \right] \Theta \left( k_{\text{F}}^{\tau}\left( r \right) -k \right) \\
		&+8\left\{ \kappa _{\text{dir}}Y\left( k,2 \right) -\left[ Y\left( k,4 \right) \right] ^2 \right\},
	\end{aligned}\label{eq:22}
\end{align}
where
\begin{align}
	c_{\mu}^{-1}Y\left( k,\mu \right) =\!\frac{\text{e}^{-\tilde{k}_{+}^{2}}-\!\text{e}^{-\tilde{k}_{-}^{2}}}{2\tilde{k}}\!+\!\int_0^{\tilde{k}_+}{\text{e}}^{-y^2}\text{d}y+sgn\left(\tilde{k}_- \right)\!\int_0^{\left|\tilde{k}_- \right|}{\text{e}}^{-y^2}\text{d}y.\label{eq:23}
\end{align}
Choosing the Gaussian correlation function
\begin{align}
	f\left( r \right) =1-\text{e}^{-\alpha ^2r^2}\label{eq:24}
\end{align}
with
\begin{align}
	\begin{aligned}
		c_{\mu}&=\frac{1}{8\sqrt{\pi}}\left( \frac{\mu}{2} \right) ^{3/2},& \tilde{k}=\frac{k}{\alpha \sqrt{\mu}},\\
		\tilde{k}_{\pm}&=\frac{k_{\text{F}}^{\tau}\left( r \right) \pm k}{\alpha \sqrt{\mu}},& sgn\left( x \right)=\frac{x}{|x|},
	\end{aligned}\label{eq:25}
\end{align}
the direct part of the Jastrow wound parameter $ \kappa_{\text{dir}} $ can be written as
\begin{align}
	\begin{aligned}
		\kappa_{\text{dir}}&=\frac{2k_{\text{F}}^{\tau}\left( r \right) ^3}{3\pi ^2}\int{\left[f\left( r \right) -1 \right]}^2\text{d}r\\
		&=\frac{1}{3\sqrt{2\pi}}\left( \frac{k_{\text{F}}^{\tau}\left( r \right)}{\alpha} \right) ^3,
	\end{aligned}\label{eq:26}
\end{align}
which is a rough measure of the importance of $ NN $ correlations and  the convergence rate of cluster expansions in the LOC approximation. From  \cref{eq:26}, $ \kappa_{\text{dir}} $ can be changed by adjusting the corresponding parameter $\alpha$ in $ f(r) $. In the calculated frameworks of LOC approximation, a large $ \kappa_{\text{dir}} $ implies a strong correlation, whereas $ \kappa_{\text{dir}}=0 $ signifies the disappearance of the $ NN $ correlation. Based on the nuclear matter calculation, a reasonable range  corresponding to the parameter $1.1 \ \text{fm}^{-1}< \alpha  < 1.7 \ \text{fm}^{-1}$ is $  0.1<\kappa_{\text{dir}}<0.3 $ \cite{Flynn1984,Stringari1990}.

\section{Result}\label{sec:3}
\subsection{NMD from the deformed RMF model}\label{sec:3.1}

As discussed in Refs.  \cite{Deguerra1991,Sarriguren2007}, the most important component of the NMDs is the monopole component of $n\left( k_{\bot},k_z \right) $. Hence, in this subsection, we discuss the results of monopole parts of $n\left( k_{\bot},k_z \right) $ for the deformed nuclei from the axially deformed RMF model. For convenience, the monopole component of total NMD is represented by $n\left( k \right) $.\par
 \begin{figure}[H]
	\centering
	\includegraphics[scale=1.0]{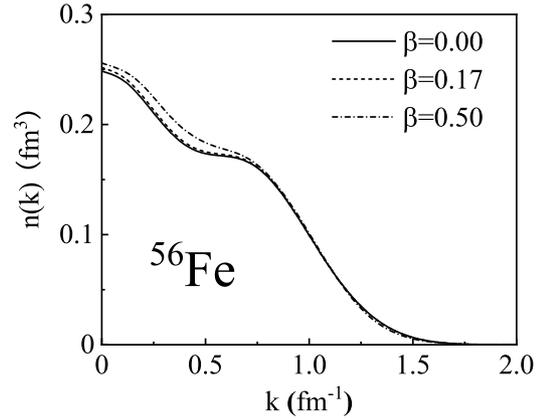}
	\caption{Total NMD $n\left( k \right) $ of $ ^{56}\text{Fe} $ with the spherical shape $ \left( \beta=0.00 \right) $ and the deformed shapes $  \left(  \beta=0.17 \ \text{and}   \ \beta=0.50 \right) $. The normalization is $ \int n\left( k\right) \text{d}\vec{k}=1$. }
	\label{fig:1}
\end{figure}
 First, we discuss the differences and similarities of  $n\left( k \right) $ calculated using different deformation parameters $ \beta $ for  $ ^{56}\text{Fe} $. It is known that the ground state of $ ^{56}\text{Fe} $ is deformed to a certain extent. In the axially deformed RMF model, we found that the binding energy is lowest at $ \beta=0.17 $. Thus, we provide the $n\left( k \right) $ of $ ^{56}\text{Fe} $ with the spherical shape $ \left( \beta=0 \right) $ and  deformed shape $\left(  \beta=0.17\right) $ in \cref{fig:1}, which are calculated using the RMF model with $ \text{NL3*} $ parameter set. To reveal the influence of large deformations, we also present NMD for the configuration $\left(  \beta=0.50\right) $ in \cref{fig:1}.  For the  $n\left( k \right) $ of $ ^{56}\text{Fe} $, the changing trends of different shapes agree with each other, and there are visible differences for  $n\left( k \right) $ in the low-momentum region $   (k<0.75 \ \text{fm}^{-1})  $.\par
\begin{figure*}[ht]
	\centering
	\includegraphics[scale=0.8]{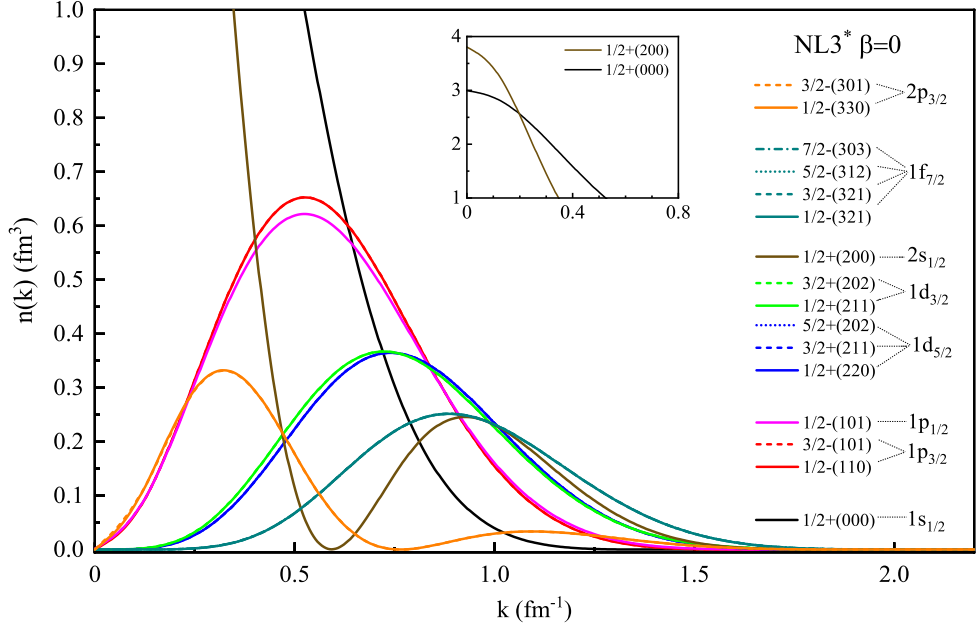}
	\caption{single-particle NMD $ n_{\varOmega}\left( k\right)  $  of the states  $ \varOmega^{\pi} $  split from the spherical $ \left( nlj\right)   $ orbitals of $ ^{56}\text{Fe} $ for the $ \beta=0 $ configuration. For one group of $ n_{\varOmega}\left( k\right)  $  split from the same spherical $  \left( nlj\right)  $ orbital, different types of lines with the same color are used. The normalization is $ \int n_{\varOmega}\left( k\right) \text{d}\vec{k}=C$, where $ C $ represents the  occupation number of the corresponding state. }
	\label{fig:2}
\end{figure*}
\begin{figure*}
	\centering
	\includegraphics[scale=0.8]{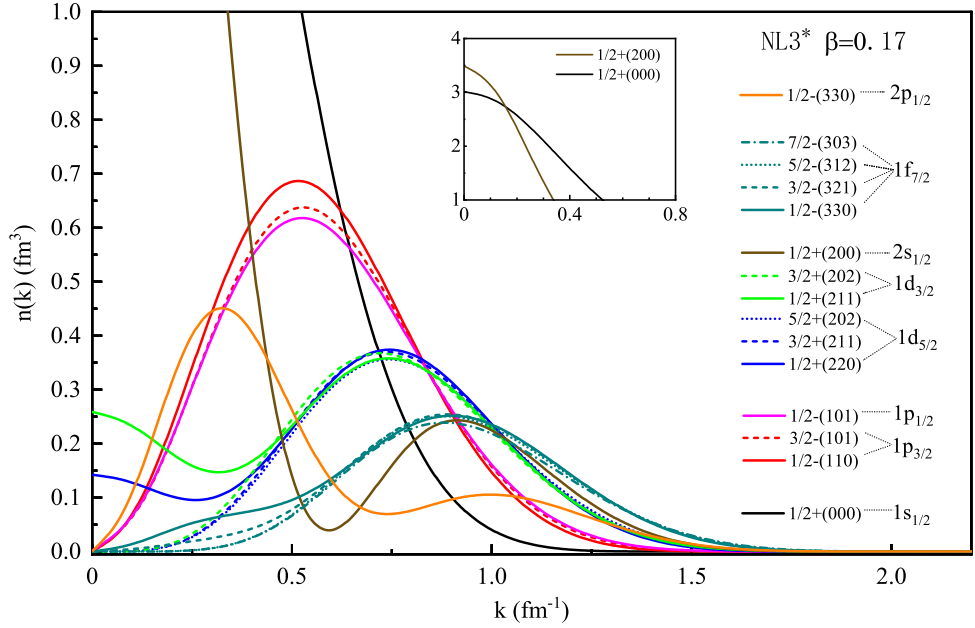}
	\caption{The same as in \cref{fig:2} but $  \beta=0.17. $}
	\label{fig:3}
\end{figure*}

To further analyze the differences and investigate the compositions of $n\left( k \right) $ in $ ^{56}\text{Fe}$, we calculate all contributions from the single-particle orbits using \cref{eq:10}. In the axially symmetric deformed shape, the rotational symmetry is broken and the condition of $ \left( 2j+1\right)  $ degeneracy is not met. However, the projection of $ j $ on the symmetry axis $\varOmega$  and the parity $ \pi $ are still good quantum numbers. Thus, the states  $ \varOmega^{\pi}\left( Nn_z\Lambda\right)  $ are used to describe the components of the spherical $ \left( nlj\right)  $ orbitals. With the axially deformed RMF model, the single-particle NMDs of states $ \varOmega^{\pi} $  of $ ^{56}\text{Fe} $ are investigated for spherical and deformed configurations, respectively; the results are presented in \cref{fig:2,fig:3}. For convenience, the single-particle NMDs of states $ \varOmega^{\pi} $  are represented by $ n_{\varOmega}\left( k\right)  $. \par
For the spherical case $ \beta=0 $ in \cref{fig:2}, $  \left( nlj \right)  $ orbits also split into $ \left( 2j+1\right)/2  $  projected states  $ \varOmega^{\pi}\left( Nn_z\Lambda\right)  $  within the theoretical framework of the axial symmetry. However, the nucleus still has rotational symmetry at this configuration. Notably, the single-particle NMDs of states split from the same spherical $ (nlj) $ orbital exactly coincide with each other  in \cref{fig:2}. Thus, in momentum space, the wave functions of the states $ \varOmega^{\pi} $  from the same $ \left( nlj \right)  $ orbital are degenerate for the spherical configurations within the axially deformed RMF model. There are also some features for $ n_{\varOmega}\left( k\right)  $   of deformed single-particle states $ \varOmega^{\pi} $  in \cref{fig:2}: (i) at $  k=0\ \text{fm}^{-1} $,  $ n_{\varOmega}\left( k\right)  $ equals 0 if the angular momentum $ l>0 $; (ii) the node of $ n_{\varOmega}\left( k\right)  $  equals $ n-1 $, where $ n $ is the principal quantum number; (iii) as  $ l $ increases, the   $ n_{\varOmega}\left( k\right)  $  moves toward the high momentum, and peaks values gradually decrease.\par
According to  $ n_{\varOmega}\left( k\right)  $  in \cref{fig:2}, the behavior of the total NMD  $n\left( k \right) $ of $ ^{56}\text{Fe}$ in \cref{fig:1} can be explained. At $ k< 0.5 \ \text{fm}^{-1} $, the  $n\left( k \right) $ is dominated by two $  1/2+ $ states $  \left( 000\right)  $ and $ \left( 200\right)  $, which cause  $n\left( k \right) $ in \cref{fig:1} to be flat first and then decline rapidly. In $ 0.5\ \text{fm}^{-1} < k < 0.8\ \text{fm}^{-1} $, the contributions of the remaining states focus on this region, which makes the drop of  $n\left( k \right) $ in \cref{fig:1} slow. For $ k>0.8 \ \text{fm}^{-1} $, the contributions of all states rapidly decrease, which also makes the trends of $n\left( k \right) $ decline rapidly in \cref{fig:1}.\par
In \cref{fig:3} we also present the $ n_{\varOmega}\left( k\right)  $ of $ ^{56}\text{Fe}$ in the deformed case, which illustrates the influences of nuclear deformations on single-particle states in momentum space compared with the results in \cref{fig:2}. Owing to the breaking of rotational symmetry, the degeneracy is removed for the axially deformed configuration, and the states divided from the same orbit deviate from each other, such as the $ 1\text{f}_{7/2} $ orbit. An interesting phenomenon is found because the deformation  values of $ n_{\varOmega}\left( k\right)  $  significantly increase at the positions as follow: (i) the nodes of all states, such as $ k=0.6 \ \text{fm}^{-1}$ for $ 1/2+(200) $ state of $ 2\text{s}_{1/2} $ and $ k= 0.7 \ \text{fm}^{-1}$ for $ 1/2-\left( 330\right)  $ state of $ 2\text{p}_{1/2} $; (ii) $ k=0 \ \text{fm}^{-1} $ for the $ 1/2+ $ states with $ l>1 $ , such as $ 1/2+\left( 220\right)  $ state of $ 1\text{d}_{3/2} $ and $ 1/2+\left( 211\right)  $ state of $ 1\text{d}_{5/2} $. This implies that the deformation produces a strong bump at these positions, which causes the $  n\left( k\right)  $  in \cref{fig:1} for the deformed case to be higher than the spherical case in the low-momentum region. However, the magnitudes and locations of all the curve peaks are almost constant, which explains why the $ n(k) $ in \cref{fig:1} for different cases exhibit the same trends. Through contrastive analyses, we can identify the contributions of each orbital for $ ^{56}\text{Fe}$ and the variation laws of degeneracy breaking in the momentum space.\par

\subsection{Relativistic effects on the NMD}
\begin{figure}[H]
	\centering
	\includegraphics[scale=1.0]{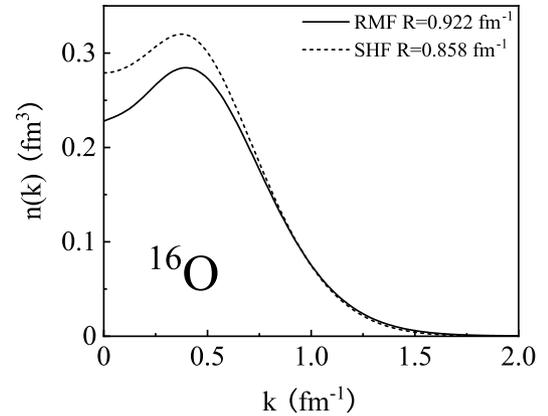}
	\caption{Total proton momentum distributions of $ ^{16}\text{O}$ from the RMF and SHF models. The RMS radii $ R $ of $ n\left( k\right) $ in momentum space are also shown. The normalization is $ \int n\left( k\right) \text{d}\vec{k}=1$. }
	\label{fig:add1}
\end{figure}

In Ref.  \cite{Gaidarov2009}, the properties of the NMD were systematically investigated using the deformed SHF model. In this subsection, we compare the total and single-particle NMDs between the RMF and SHF models. We choose $ ^{16}\text{O}$ as the candidate nucleus because it has only three orbits, which is convenient to analyze the relativistic effect on NMD. For the RMF model, the $ \text{NL3*} $ parameter set is used in the calculations. For the SHF model, we calculate the NMD using the code HFBTHO (v1.66p) \cite{HFBTHO}  with the SKP parameter set. The total proton momentum distributions $ n\left( k\right)  $ from the RMF and SHF models are calculated; they are presented in \cref{fig:add1}, where both results are normalized to 1. The tendencies of the $ n\left( k\right)  $ are markedly similar in the two models. However, some notable differences exist in  $ n\left( k\right)  $. The  results from the RMF model are significantly smaller than those from the SHF model in the region  $ k<0.75 \ \text{fm}^{-1} $ and shift outward at  $ k>1.00 \ \text{fm}^{-1} $. This causes  the distribution from the RMF model to have a larger root mean square (RMS) radius in momentum space than that from the SHF model.\par
\begin{figure}[H]
	\centering
	\includegraphics[scale=1.0]{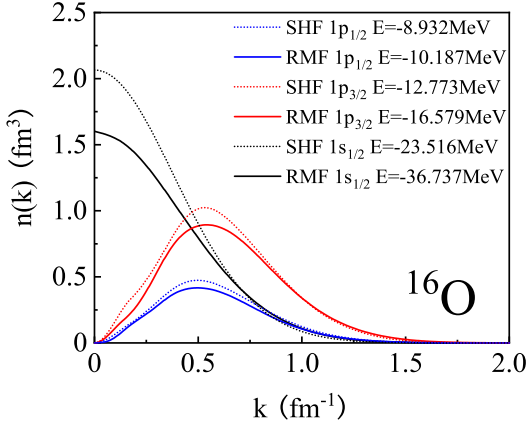}
	\caption{single-particle NMD of $ ^{16}\text{O}$ from the RMF and SHF models. The orbital energies are also shown. The normalization is $ \int n\left( k\right) \text{d}\vec{k}=C$,where $ C $ denotes the  occupation numbers of corresponding orbits.}
	\label{fig:add2}
\end{figure}

To explain the differences in the total proton momentum distributions between the RMF and SHF models in \cref{fig:add1}, we further analyze the single-particle NMDs from these two models in \cref{fig:add2}, where all results are normalized to the  occupation numbers of corresponding orbits. From \cref{fig:add2}, the orbital energies from the RMF model are systematically larger than those from the SHF model. All single-particle NMDs from the RMF model have a lower distribution in the region of $ k<0.75 \ \text{fm}^{-1} $ and diffuse outward in the region of $ k>1.00 \ \text{fm}^{-1} $. Moreover, \cref{fig:add2} shows that the relativistic effects on different orbits are related to the order of energy levels. Compared with p orbits, larger differences exist in momentum distributions and orbital energies for s orbits between the RMF and SHF models. With the increase of energy level, the relativistic effects gradually decrease, which makes the results of the RMF and SHF models close to each other.\par
Combining \cref{fig:add1} and \cref{fig:add2}, we can discuss the relativistic effects on the total and single-particle NMDs. The self-consistent central potentials from the RMF model are deeper than those from the SHF model, which causes the energies of different orbits of the RMF model to be larger than those of the SHF model \cite{Ebran2012}. In the mean-field model, the average momentum of certain orbit is closely connected with the orbital energy. For certain orbits, the increase in orbital energy increase of RMS radius in momentum space. Therefore, the single-particle NMDs from the RMF model have central depressions and outward shift. Similar results were also obtained in coordinate space. In Figure 3 of Ref.  \cite{Pannert1987}, the deeper potential of the RMF model causes the inward contraction of the wave function, which increases the center densities and reduces the RMS radii. This corresponds with the results in momentum space (\cref{fig:add1}). Based on the comparisons of the total and single-particle NMDs between the RMF and SHF models, the relativistic effects on the single-particle wave functions of momentum space can be reflected.  \par
\subsection{Effects of $ NN $ correlations on NMD}
In this subsection, we discuss the effects of $ NN $ correlations in $^{56}\text{Fe}$. Based on the RMF model, the $ NN $ correlations are considered by the LFD  and LDA methods of \cref{sec:2.2} to correct the high-momentum components.  In the framework of the LFD method, because the particle-state contributions to NMD in its natural orbital representation are almost equal for all nuclei, the $ NN $ correlations for finite nuclei are described by the rescaling of the one for deuteron. In the framework of the LDA approach, supposing that the $ NN $ correlations are almost unaffected by surface and shell effects, the correlations for finite nuclei are obtained from the calculations of uniform nuclear matter at different densities using the LOC approximation.\par
\begin{figure}[H]
	\centering
	\includegraphics[scale=1.0]{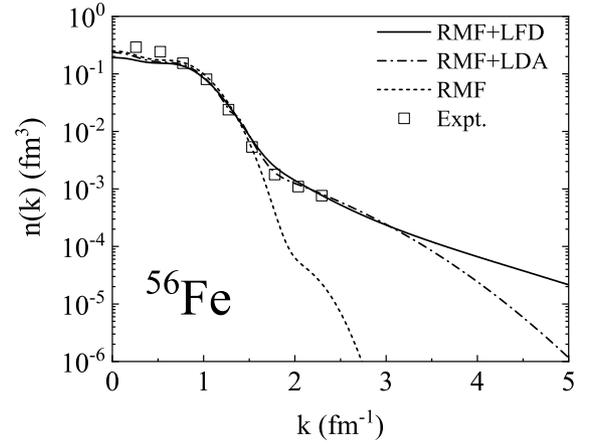}
	\caption{ Total NMDs of $ ^{56}\text{Fe} $ with the $ NN $ correlations. The deformed parameter $ \beta=0.17 $ for different $ NN $ correlations. The data on the NMD $ n\left( k\right)  $ extracted by the nuclear $y$-scaling analysis are from Ref.  \cite{ciofi1991}.}
	\label{fig:4}
\end{figure}

The total $ n\left( k\right)  $ of $ ^{56}\text{Fe} $ with and without the $ NN $ correlations are shown in \cref{fig:4}. In the RMF+LFD method, we consider the scaling factor $ C_{A} $ as 4.5 \cite{ciofi1996}, whereas, in the RMF+LDA method, we choose the value of parameter $ \alpha=1.1 \ \text{fm}^{-1} $ \cite{Stringari1990}. For verifying the validity of theoretical studies, the $ n\left( k\right)  $ extracted from $y$-scaling  analyses on the quasielastic electron scattering experiments \cite{ciofi1991,ciofi1996} are also employed for comparison. Because of the lack of $ NN $ correlations, the results of the pure RMF calculation deviate from the experimental data by more than $  k>1.5\ \text{fm}^{-1} $. In contrast, the value of $ n\left( k\right)  $ on the tail increases in the RMF+LFD and RMF+LDA methods by considering the $ NN $ correlations, which coincides with experimental data. The results show that high-momentum components of $ n\left( k\right)  $ in $^{56}\text{Fe}$ are related to the $ NN $ correlations. However, owing to the differences in the physical models of the LFD and LDA methods, some differences also exist between these two results in \cref{fig:4}. Therefore, in the following, we perform a fine-grained analysis for the $ NN $ correlations of the LFD and LDA methods separately.\par
\begin{figure}[H]
	\centering
	\includegraphics[scale=1.0]{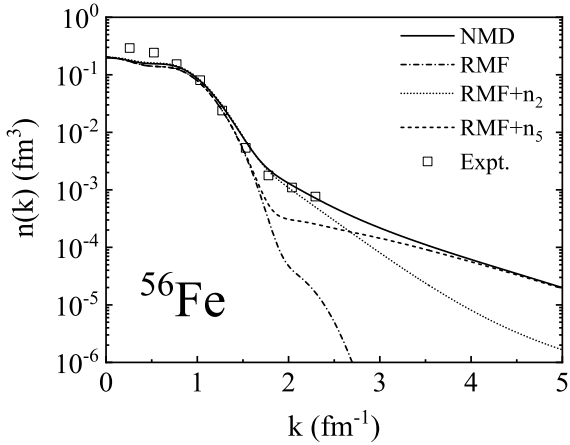}
	\caption{Momentum distribution of $ NN $ correlations components in $ ^{56}\text{Fe} $. ``NMD" denotes the total NMD. ``RMF", ``$ n_{2} $", and ``$ n_{5} $"  denote NMDs contributed by corresponding component defined in \cref{eq:19}.}
	\label{fig:5}
\end{figure}

For the LFD method, we  explore the physical origins of the correlations in  $ ^{56}\text{Fe}$ by adding the contributions of $ n_{2} $ and $ n_{5} $ separately in \cref{fig:5} with the values extracted from experimental data. As can be seen in \cref{fig:5}, there are significant differences for two sets of NMD by adding $ n_{2} $ and $ n_{5} $ terms. The results of the $\text{RMF}+ n_{2} $ component nicely reproduce the experimental values in the region $ 0<k<2.5 \ \text{fm}^{-1} $. It must also be mentioned that the contribution of $ n_{2} $ is obtained from  $ f_2 $, which coincided with the Bonn D-wave \cite{Carbonell1995}, therefore the $\text{RMF}+ n_{2} $ component in \cref{fig:5} can be attributed to the correlation induced from the tensor force. For the results of $\text{RMF}+ n_{5} $, no tensor is included in this component. Hence, we observe a deep valley structure around $ k \approx 2 \ \text{fm}^{-1} $, which significantly deviates from the experimental values. This comparison demonstrates the influences of the tensor correlation on the  $ n\left( k\right)  $ of heavy nuclei.\par
\begin{figure}[H]
	\centering
	\subfigure{
		\begin{minipage}{8cm}
			\centering	
			\includegraphics[scale=1.0]{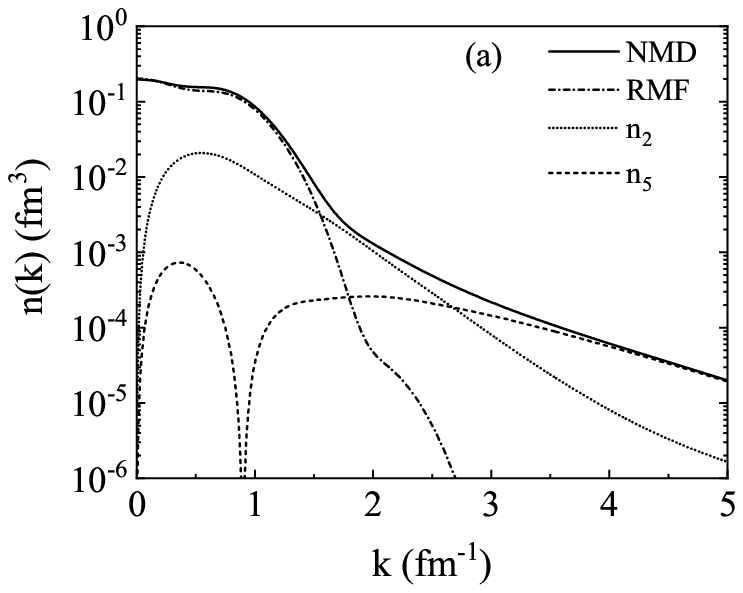}
		\end{minipage}\label{fig:6a}	
	}
	\subfigure{
		\begin{minipage}{8cm}
			\centering	
			\includegraphics[scale=1.0]{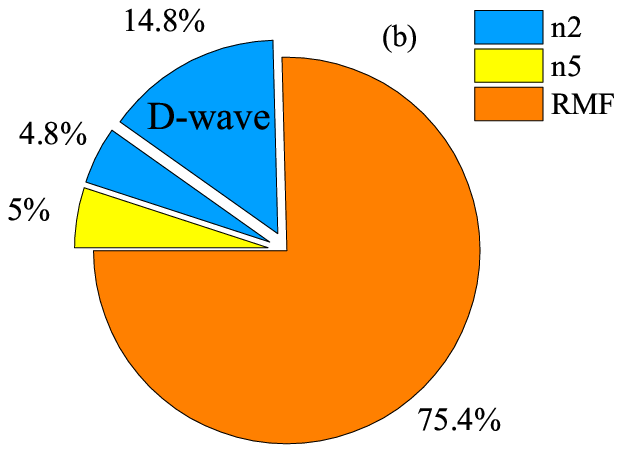}
		\end{minipage}\label{fig:6b}
	}
	\caption{(a) Decomposition of the high momentum component of  $ ^{56}\text{Fe} $. The definitions of the corresponding components are the same as those in \cref{fig:5}. (b) Probabilities of each  component of $ n_{2} $, $ n_{5} $, and RMF in $ ^{56}\text{Fe} $. The contribution of D-wave is also considered.}
\end{figure}

Further, we decompose each component of $n\left( k \right) $ of  $ ^{56}\text{Fe}$ in the LFD model based on \cref{eq:12,eq:18} of \cref{sec:2}, and the results are presented in \cref{fig:6a}. In this figure, the contributions from RMF, $ n_{2} $, and $ n_{5} $ terms are shown together for comparison. In the low-momentum region $  k<2.0\ \text{fm}^{-1} $, the mean-field term provides a good description for $n\left( k \right) $ in heavy nuclei. However, the high-momentum components of $n\left( k \right) $ in the region $  k>2.0\ \text{fm}^{-1} $ can not be explained without the contributions of the $ n_{2} $ and $ n_{5} $ terms. As discussed above, the $ n_2 $ term is mainly induced by the tensor force and dominates around $ k \approx 2 \ \text{fm}^{-1} $. The $ n_5 $ term is mainly induced by the $ \pi$-exchange \cite{Carbonell1995} and dominates around $ k \approx 4 \ \text{fm}^{-1} $. \par
To obtain the quantitative analyses of the contributions of each term, we calculate the probabilities of each term and present the results in \cref{fig:6b}. Based on the data, the $ NN $ correlations contribute to   24.6\%  of the total NMD in  $ ^{56}\text{Fe}$, which is comparable with the estimation of 20\% in Ref. \cite{ohen2014} and 23\% in Ref.  \cite{Lyu2020} even though different nuclei are discussed in these works. The contribution of the $ n_2 $ term is 19.6\%, in which the contribution of the D-wave component is 14.8\%. This result is consistent well with the 13\% for D-wave components in Ref.  \cite{Myo2017}. The contribution of $ n_2 $ is 4.8\% greater than that of the D-wave because the $ n_2 $ term  contains few the S-wave components \cite{Carbonell1995}. Thus, it can be concluded that the correlations induced by the tensor force play the dominant role in the region  $ 1.5 \ \text{fm}^{-1}<k<2.5 \ \text{fm}^{-1} $. In contrast, the $ n_5 $ term dominates the region $  k>3.0\ \text{fm}^{-1} $ although their 5\% probability  is comparably small. \par
\begin{figure}[H]
	\centering
	\includegraphics[scale=1.0]{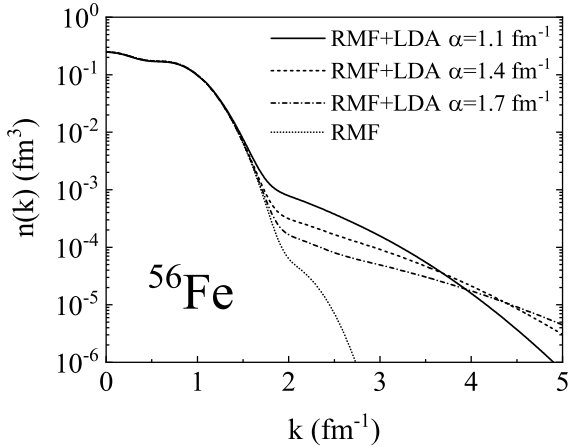}
	\caption{Influence of different $ NN $ correlation strengths on NMD in the RMF+LDA method. The deformed parameter $\beta$ for $ ^{56} \text{Fe} $ is 0.17.}
	\label{fig:7}
\end{figure}

In the RMF+LDA method, the information about the $ NN $ correlation in   $ ^{56}\text{Fe}$  is mainly embodied in the correlation function $ f(r) $ of \cref{eq:25}. Based on microscopic nuclear matter calculations, the correlation parameter $\alpha$ is set to $ 1.1\ \text{fm}^{-1} $, which can reproduce the momentum distributions $n\left( k \right) $ in finite nuclei. For comparison, we also choose two more values  $ \alpha=1.4\ \text{fm}^{-1}$ and $ \alpha=1.7 \ \text{fm}^{-1}$ to analyze the impacts of $ NN $ correlation strength on $n\left( k \right) $. From \cref{fig:7}, with the increase in  $\alpha$, $n\left( k \right) $ experience a decline in the region $ 1.5 \ \text{fm}^{-1}<k<2.5 \ \text{fm}^{-1} $. In previous researches \cite{Alvioli2013,Neff2015}, the $n\left( k \right) $ in this region is dominated by the effects of the tensor force. From \cref{eq:26}, the Jastrow wound parameter $ \kappa_{\text{dir}} $ gradually drops with increasing $ \alpha $, implying that the correlation effects also weaken overall. The results of $n\left( k \right) $ in \cref{fig:7} confirm this conclusion.\par
\begin{figure}[H]
	\centering
	\includegraphics[scale=1.0]{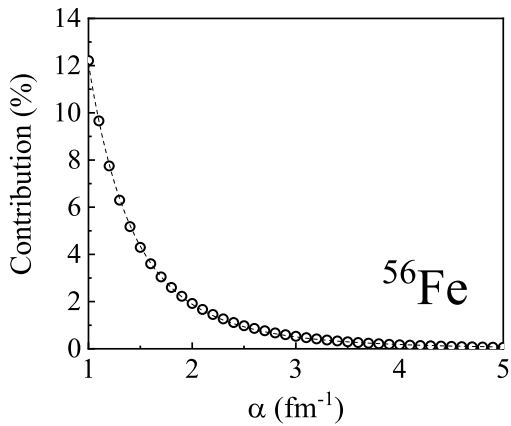}
	\caption{Changes in the contributions of $ NN $ correlations above $ k_{\text{F}} $ with parameter $\alpha$ in $ ^{56}\text{Fe}$.}
	\label{fig:8}
\end{figure}

For the RMF+LDA method, the proportions of $n\left( k \right) $ above Fermi momentum $ k_F $ are calculated for different correlation parameters $ \alpha $ from $ 1.0\ \text{fm}^{-1} $ to $ 5.0\ \text{fm}^{-1} $, and the results are shown in \cref{fig:8}. The proportions exponentially decrease with the increase of  $\alpha$. For $ \alpha=1.1\  \text{fm}^{-1}  $, the $ NN $ correlation contributes to $ 9.6\% $ of the total $n\left( k \right) $, which is smaller than the probability of contribution induced by tensor force in previous studies \cite{Myo2017}. Moreover, in \cref{fig:7}, similar valley structures as those shown in \cref{fig:5} emerge in the region dominated by the tensor force with increasing  $\alpha$. However, how to correctly reflect the tensor contributions in the framework of the LDA method is still an open problem.

\section{Conclusion}\label{sec:4}
Based on previous studies, we deduce a framework for calculating NMDs in heavy nuclei based on the axially deformed RMF model, with $^{56}\text{Fe}$ as the representative nucleus in the calculations.  The total and single-particle NMDs from the RMF model are discussed in the spherical and deformed cases, respectively. Comparing the total NMDs for different configurations, one can observe the effects of the deformation on the nuclear structure from the RMF calculations in momentum space. For the single-particle orbits,  the deformation  can lead to the breaking of the $ \left( 2j+1\right) $ degeneracy in spherical symmetry, where each spherical orbit $  \left( nlj \right)  $ splits into $ \left( 2j+1\right) /2 $ different deformed states $ \varOmega^{\pi} $.\par 
By comparing the NMDs from the RMF and SHF models, we analyze the relativistic effects on the NMD. The RMF model generates deeper self-consistent central potentials than the SHF model, which increases the energies of single-particle orbits. For certain orbits, the increase in orbital energy increases the RMS radius in momentum distribution, which appears as central depressions and outward shift. These results agree with the results in coordinate space in previous studies.\par
The effects of the $ NN $ correlations are further introduced into the NMD  of the deformed RMF model via the LFD and LDA methods. The high-momentum region of NMD is modified by the $ NN $ correlations, which is consistent with the experimental data. Moreover, we provide in-depth analyses for the different types of $ NN $ correlations in these two methods. For the RMF+LFD method, by  decomposing the total NMD into the components of RMF contribution and different correlation terms, it is found that the $ NN $ correlations contribute to  24.6\%  of the total NMD in which 14.8\%  is induced by the tensor force. Further \cref{fig:5} shows that the $ NN $ correlations induced by the tensor force dominate the range  $ k\approx 2\ \text{fm}^{-1} $. For the RMF+ LDA method in which a Jastrow-type correlation function is adopted, with the increase in the correlation parameter $ \alpha $, $ n\left( k\right)  $ in the region dominated by the tensor force  ($ k\approx 2 \ \text{fm}^{-1} $) decreases. Further data analyses demonstrate that the total contributions of correlation above the Fermi momentum $ k_{\text{F}} $ exponentially decrease  with increasing $\alpha$. Based on the analyses of the behaviors and contributions of $ NN $ correlations, the effects of the tensor force on NMD in heavy nuclei are clarified.\par
Herein, we conduct a preliminary study on NMDs in heavy nuclei based on the framework of the RMF model, and positively demonstrates the validity of this model  in momentum space. In the future, we will continue to study the NMD  via the microscopic beyond mean-field theory. This study is expected to be useful for theoretical and experimental studies on the nuclear structures and  reactions.

%%%%%%%%%%%%%%%%%%%%%%%%%%%%%%%%%%%%%%%%%%%%%%%%%%%%%%%
%%% Acknowledgements. ???
%%%%%%%%%%%%%%%%%%%%%%%%%%%%%%%%%%%%%%%%%%%%%%%%%%%%%%%
\Acknowledgements{This work was supported by the National Natural Science Foundation of China (Grants No. 11505292,  No. 11605105, No. 11822503, No. 11975167, and No. 12035011), by the Shandong Provincial Natural Science Foundation, China (Grant No. ZR2020MA096), by the Fundamental Research Funds for the Central Universities (Grant No. 20CX05013A, No. 22120210138), and by the Graduate Innovative Research Funds of China University of Petroleum (East China) (Grant No. YCX2020104).}

\bibliography{references}
\end{multicols}
\end{document}